\documentclass[prl,showpacs,twocolumn]{revtex4}
\usepackage{graphicx}
\usepackage{dcolumn}
\usepackage{bm}
\begin{document}
\title{Tuning surface metallicity and ferromagnetism
by hydrogen adsorption at the polar ZnO(0001) surface}

\author{N. Sanchez, S. Gallego, J. Cerd\'a and M.C. Mu\~noz} 
\email{mcarmen@icmm.csic.es}
\affiliation{
Instituto de Ciencia de Materiales de Madrid, Consejo Superior de
Investigaciones Cient{\'{\i}}ficas, Cantoblanco, 28049 Madrid, Spain}

\date{\today}
\pacs{71.20.Tx, 73.22.-f, 71.70.Ej}

\begin{abstract}

The adsorption of hydrogen on the polar Zn-ended ZnO(0001) surface has been 
investigated by density functional {\it ab-initio} calculations. An 
on top H($1 \times 1$) ordered overlayer with genuine H-Zn chemical bonds is 
shown to be
energetically favorable. The H covered surface is metallic and spin-polarized,
with a noticeable magnetic moment at the surface region. 
Lower hydrogen coverages lead to strengthening of the H-Zn bonds, corrugation 
of the surface layer and to an insulating surface. 
Our results explain experimental observations of hydrogen adsorption on this surface, 
and not only predict a metal-insulator transition, 
but primarily provide a method to reversible switch surface magnetism
by varying the hydrogen density on the surface.
\end{abstract}

\maketitle

ZnO is one of the most technologically important metal oxides, which holds great promise for applications
as a wide band gap semiconductor:
it shows the largest charge-carrier mobility among oxides, extraordinary catalytic properties and the 
unusual coexistence of transparency and conductivity \cite{ozgur}.
The recently reported high temperature (HT) ferromagnetism in ZnO-based thin layers 
and nanostructures \cite{ueda}
turns it additionally into a potential HT semiconducting ferromagnet, which 
could be used in magnetoelectric and magnetotransport devices
tuning simultaneously charge and spin \cite{awscha}.
However, at present, and despite numerous experimental and theoretical studies,
the mechanism behind the HT magnetic order in ZnO is still under debate \cite{chambers,prl,sundar,xu}.

Particular attention deserves the role of H doping in ZnO.
Hydrogen is a very reactive element that exhibits qualitatively different 
behaviour in different media.
It is amphoteric, can act either as donor (H+) or acceptor (H-), 
can occupy different lattice sites and even modify the host structure,
and in general counteracts the conductivity of the host.
Isolated hydrogen has been found to act as a shallow donor in bulk ZnO, and thus 
it has been attributed as a source of the unintentional n-type conductivity exhibited by ZnO \cite{vwalle}.  

At surfaces, the interaction with ambient H is almost unavoidable even in ultra-high-vacuum conditions.
Moreover, the presence of hydrogen has a pronounced
influence which can even change the surface electronic properties.
Understanding the interaction of hydrogen with the ZnO surfaces
is crucial in order to control the 
properties of ZnO-based low-dimensional structures.
In recent years, systematic investigations of the
different ZnO low-indexed surfaces have been performed \cite{woll}. There is agreement
in the formation of an ordered hydrogen overlayer on both 
the non-polar ZnO ($10\overline{1}$0) and the polar O-ended ZnO
($000\overline{1}$) surfaces.
Much few work has been devoted to the Zn-terminated ZnO(0001) surface, where
the interaction of H atoms is thought to be the weakest among the 
ZnO surfaces, since the binding energy of Zn–H pairs should be considerably weaker
than that of O–H bonds.  
An experimental study revealed that,
exposing this surface to atomic hydrogen, 
an ordered ($1 \times 1$) overlayer consisting of Zn-hydride species is formed \cite{becker}.
However,
larger exposures to both atomic and molecular hydrogen cause the unexpected
unstability of the H ($1 \times 1$) overlayer, leading
to a complete loss of lateral order which reflects a 
random distribution of H adatoms.



The aim of this Letter is to show that the adsorption of atomic hydrogen on the polar 
Zn-ZnO(0001) surface gives rise to an ontop H($1 \times 1$) ordered overlayer
with genuine H-Zn chemical bonds. 
As shown below, surface magnetism and metallicity are distinct 
characteristics of this two-dimensional (2D) H($1 \times 1$) ordered overlayer.
Furthermore, the partial coverage of the surface with H leads to reinforcement 
of the H-Zn bonds, corrugation of the surface layer and, more interestingly,
the emergence of a spin-paired insulating state. Hence, a metal-insulator 
transition accompanied by the extinction of the magnetization 
can be driven by reducing the H density on the surface from 1 ML down to 1/2 ML.
Further decrease of the H coverage restores the surface metallicity.
 
Our calculations are based on density functional theory employing
norm-conserving pseudopotentials and localized numerical atomic orbitals (AO) 
as implemented in the SIESTA code\cite{siesta}. We consider both the local 
spin density approximation (LSDA) and the generalized gradient approximation (GGA), 
with the same basis set and parameters of Ref.\cite{prl}. Hydrogen is described 
by a Double Z 1$s$ AO. In order to ensure
that our conclusions are robust against the choice of the
exchange-correlation functional, we have
performed additional total energy optimizations with both the PBE0 and
HSE hybrid functionals \cite{vasp1,hybrids} using the Projector Augmented Wave
(PAW) method implemented in the VASP code\cite{vasp}. We employed
similar parameters to those used in Ref.~\cite{kresse}, where
such functionals were sucessfully applied to study the defect energetics
in ZnO. 

The surfaces are modeled by periodically repeated slabs containing between 15 and 
17 atomic planes, separated by a vacuum region of at least 
20 $\AA$. Bulk-like behaviour is always attained at the innermost central 
layers. We calculate both symmetric and asymmetric slabs about the central plane, 
though we are always subject to the lack of inversion symmetry of the wurtzite structure.
We use ($1 \times 1$) and ($2 \times 2$) 2D-unit cells to model the surfaces with 
adsorbed hydrogen.
All the atomic positions are allowed to relax until the forces on the atoms are
less than 0.03 eV/$\AA$. 
Brillouin Zone integrations have been performed on a $12 \times 12 \times 1$ 
Monkhorst-Pack supercell ($12 \times 12 \times 8$ for bulk structures). 
Careful convergence in the k-mesh and the inclusion of relaxations are essential in order to 
accurately describe the adsorption states and magnetism.

\begin{figure}[th]
\includegraphics[width=\columnwidth,clip]{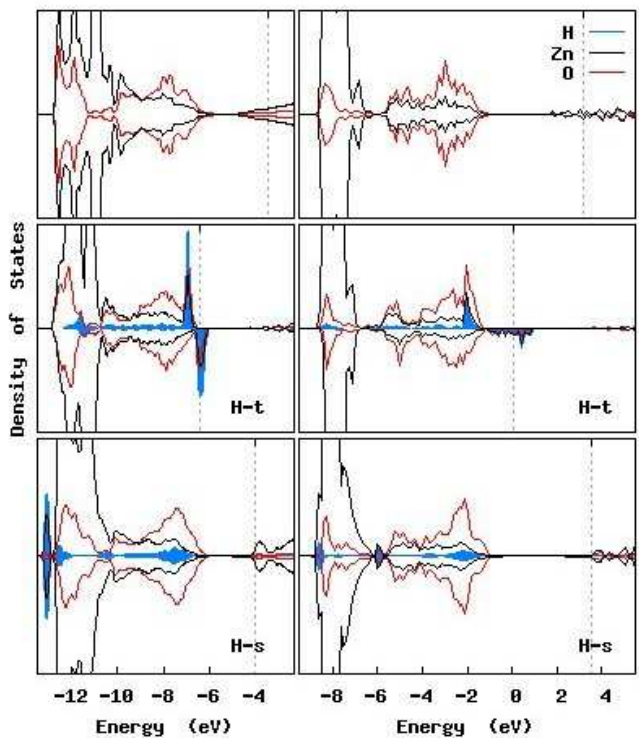}
\caption{PDOS projected on the H (filled curve) and neighbouring Zn and O atoms of the (top) 
bare Zn-Zn(0001) surface, (middle) 1 ML H covered surface, and (bottom) bulk ZnO
with a H impurity in the MBC. Left panels correspond to LDA, right ones to HSE.}
\label{dos}
\end{figure}

For 1 ML coverage, the stable adsorption site of H was determined to be atop the Zn atoms
after exhaustive minimization considering several surface and subsurface positions, 
including fcc and hcp hollow, bridge and off-symmetric sites.
We present in Figure~\ref{dos} the density of states projected (PDOS) onto 
the outermost layers for the bare and H covered surface, both using 
LDA and the HSE functional. For comparison, the PDOS corresponding to a 
substitutional H atom at an oxygen site in bulk ZnO -multicenter bonds 
configuration (MBC) \cite{vwalle}- is also shown.
As expected, the HSE functional provides a better description of the band
gap and increases the localization of the Zn $d$-states.
The calculated gaps using LDA and HSE for bulk ZnO are
0.80 and 3.45~eV, respectively, compared to the experimental value of
3.60~eV. Nevertheless, both calculations provide the same physical picture.
The clean surface exhibits the well-known metallic character, with
surface states (SS) at the bottom of the conduction band (CB) \cite{kressesurf}.
However, H adsorption creates hybridized bonding orbitals at the valence band (VB) edge, 
depleting the Zn 4$s$ SS at the CB.
The Fermi level lies in the H derived states and thus the surface is metallic and $p$-doped. 
This is opposite to the bulk MBC, for which
the H-bonding state lies deep below the VB with a total 
charge not large enough to completely deplete the Zn derived CB states. 

\begin{figure}
\includegraphics[width=\columnwidth,clip]{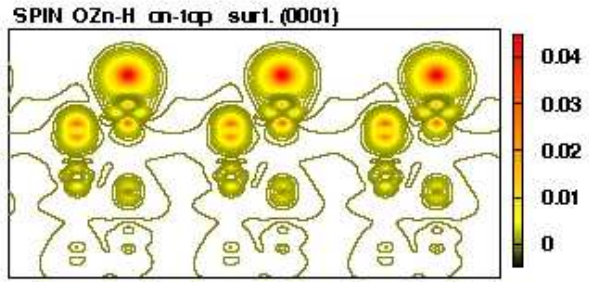}
\caption{
Side view of the spin density distribution at the ZnO(0001) surface
covered by 1 ML H.}
\label{rho}
\end{figure}

A distinct characteristic of the H covered surface in the atop geometry
is the spin-polarization of the bands.
It is not restricted to the H layer, but extends into the ZnO
subsurface leading to a net magnetization of the surface region. This 
is clearly seen in Fig. \ref{rho}.
The magnetic moments at the H, Zn and O layers are 0.29,
0.09 and 0.09 $\mu_B$, respectively, and the magnetic energy gain is
around 70 meV. Similar values are obtained in the HSE calculation, 
the total magnetization differing in less than 10\%.  
Summarizing, H adsorption at the ML coverage gives rise to a $p$-doped
metallic surface with the Fermi level pinned at the H-derived bands and
a net surface magnetic moment of around 0.5~$\mu_B$.
 

\begin{figure*}[ht]
\begin{center}
\includegraphics[scale=0.5,clip]{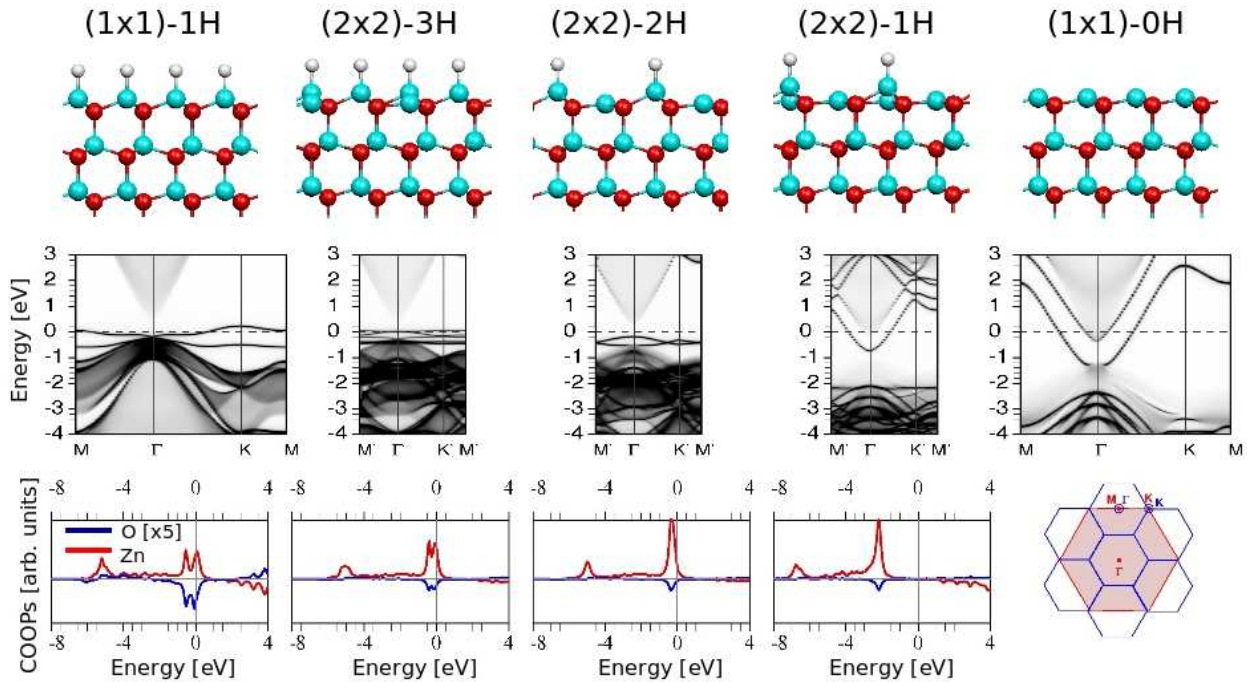}
\caption{(Top) Side views of the ZnO(0001) surface for decreasing H coverage
         from left to right. Oyxgen, Zn and H atoms are colored in red, blue 
	 and white, respectively. Below each sketch we provide for each case
         the corresponding: (middle) PDOS projected on the first surface layer 
         and resolved in $k$-space\cite{ysi2}, the BZ for the
	 (1$\times$1) and (2$\times$2) 2D cells being depicted at the right of the
	 bottom panel; (bottom) H-Zn and H-O (magnified by a factor of 5) COOPs.}
\label{ek}
\end{center}
\end{figure*}

In order to understand the emergence of the spin polarization when going from
the bare Zn surface to the complete H overlayer, we have modelled
partial Hydrogen coverages of 3/4, 1/2 and 1/4 MLs with LDA. 
The corresponding geometric and electronic structures are displayed in Figure \ref{ek} and
further details are given in Table~\ref{table}.
The adsorption of H creates low dispersion states at the top of the VB, which are spin-splitted
for 1 ML coverage (with only one spin component completely filled), and start to merge for lower
coverages until they become degenerate for 1/2 ML of H. The main result is that this loss of
spin-polarization is accompanied by a metal to insulator transition of the surface.
Thus, while at 3/4 ML the surface is still ferromagnetic and metallic, for a coverage of 1/2
ML the ground state corresponds to a non-magnetic insulator with
the Fermi level above the hydrogen derived band.
Further decrease of the hydrogen coverage results in partial
occupation of the CB states due to the increased number of unsaturated Zn dangling bonds,
restoring the surface metallicity. Therefore,
a metal to insulator transition can be tuned by varying the H-coverage
back and forth. Even more, by varying the Hydrogen density reversible switch of surface 
magnetism can be achieved. 

\begin{table}
\caption{H-related surface energy (E$_{surf}$, in eV), bond distances (d$_{AB}$, in $\AA$), 
         and Mulliken charges (Q$_A$) of the surface atoms for different H coverages
         and for the bare Zn-ZnO(0001) surface. The last column refers to undoped bulk ZnO.}
\begin{tabular}{ccccccc}
\hline \hline
          & 1 H  &    3/4 H  &   1/2 H   &   1/4 H   &  bare & bulk  \\ \hline
d$_{H-Zn}$ &1.62  & 1.60      & 1.57      & 1.56      &      &       \\
d$_{Zn-O}$ &0.71  & 0.78/0.21 & 0.91/0.20 & 1.02/0.14& 0.39 &  0.61 \\ \hline
Q$_H$     &1.09  & 1.12      & 1.17      & 1.21      &      &       \\
Q$_{Zn}$  &11.36 &11.38/11.25&11.42/11.24&11.45/11.27&11.37 & 11.23 \\ \hline
E$_{surf}$ &-1.53 &   -1.86   &   -2.51   &-3.22      &      &       \\
\hline \hline
\end{tabular}
\label{table}
\end{table}

A deeper insight about this remarkable phenomenon can be obtained regarding the nature of
the bonds -see Table~\ref{table}. 
In all cases, the H-Zn bond length is significantly shorter than at the bulk ($\sim 2$ $\AA$ in the
MBC), a hint of the formation of genuine and stronger chemical bonds
at the surface.
Furthermore, the bond length reduces as the H coverage diminishes, evidencing a reinforcement
of the H-Zn bonds.
This bond strengthening is consistent with the variation of
the Mulliken charges and the corresponding crystal overlap populations (COOPs) 
at the lower panel of Figure~\ref{ek}.
The charge at both the H and the Zn increases as the H coverage reduces, 
indicating that a larger amount of charge is shared between the two atoms. 
The COOPs confirm this scenario, their positive value corresponding to bonding states\cite{prb}, 
which are progressively filled as the H coverage reduces. 
Noticeably, Hydrogen also interacts 
with the O in the subsurface layer forming an antibonding state. 

The reinforcement of the H-Zn bonds can be understood regarding the 
atomic structures in Figure~\ref{ek} and Table~\ref{table}.
For the clean surface we find a contraction of the  
first double layer spacing, d$_{Zn-O}$, in agreement with previous calculations \cite{kressesurf}.
Contrary, the surface completely covered with H exhibits a slightly expanded
Zn-O distance. For partial H coverages the surface Zn atoms become inequivalent,
leading to two different Zn-O interlayer distances and a large corrugation
of the Zn plane: those
bonded to H experience an outward relaxation while the unbonded ones
show an inward relaxation even stronger than that at the bare surface,
becoming almost coplanar with the oxygen plane.
For deeper layers little relaxation is found.
There is a strong
correlation between the Zn-O interlayer distance and the redistribution
of charge: Zn atoms bonded to H -large d$_{Zn-O}$- show an
increase in their Mulliken charges, while those not bonded -small
 d$_{Zn-O}$- lose charge approaching the bulk values.

The energies provided in Table~\ref{table} for the H covered surfaces 
also support the formation of strong bonds with a significant covalent character.
They are calculated as the difference between the total energy of the relaxed slab 
and those of the relaxed isolated ZnO and hydrogen slabs.
Their negative values indicate that the chemisorption of hydrogen on the Zn(0001) 
surface is exothermic.
In fact, at a hydrogen density of 1/2~ML the surface energy exceeds by
140 eV the formation of the H$_2$ molecule in the gas phase, 2.37 eV/H, 
making adsorbed hydrogen highly stable against desorption.
In addition, there is a net reduction of the surface energy for all H
coverages with respect to the bare surface. This is specially significant 
for partial H coverages, where the existence of inequivalent Zn sites allows 
for the corrugation of the Zn layer and the subsequent decrease of surface
energy.
These results consistently explain experimental observations, not only about the 
formation of a H($1 \times 1$) ordered overlayer, but also the loss of surface 
order under prolongated hydrogen exposures \cite{becker}.
As shown above, partial H adsorption is energetically more favorable than the bare surface, 
which could explain the experimentally observed enhancement of the chemical 
reactivity of surfaces previously exposed to H with respect to the pristine ones.
Also the almost coplanar positions of the Zn and O atoms in the bare regions
of the partially H covered surface
can account for the oxygen signature observed in the XPS spectra \cite{becker}.

An additional important conclusion can be extracted with respect to the
ferromagnetism measured in ZnO, both undoped and 
doped with magnetic impurities. It has been observed to vary with the oxygen partial
pressure, and consequently its existence has been associated to
oxygen vacancies. However, calculations as well as careful experiments indicate
that oxygen vacancies cannot be a source of magnetism, but instead unsaturated
oxygen either due to Zn vacancies or to O-terminated surfaces\cite{prl,sundar,xu,peng}.
Our results strongly suggest that 
unintentionally adsorbed hydrogen may take active part in the
observed magnetism, particularly for undoped ZnO nanocrystals. 
Adsorbed hydrogen is sensitive to the oxygen partial pressure, since
the probability of formation of HO complexes increases with the oxygen chemical potential, 
leading to the subsequent extinction of the magnetism reported here.
Moreover, the hydrogen electron spin has been measured in In$_2$O$_3$ 
when hydrogen acts as an oxygen vacancy passivating center \cite{kumar}.

As a final remark, the Zn-ZnO(0001) surface may undertake a triangularly
shaped reconstruction under specific growth conditions \cite{dulub}.
We have investigated the effect of such reconstruction
through calculations of stepped surfaces, and the general conclusions are analogous to those 
presented here for the unreconstructed surface, although strong O-H bonds are also formed at 
step edges. 

In summary, atomic hydrogen adsorbs on the Zn-ZnO(0001)
polar surface atop the Zn atoms, forming strong H-Zn bonds
and leading to a metallic, $p$-doped surface with a net magnetic moment.
As the H coverage diminishes, there is a reinforcement of the remaining H-Zn
bonds accompanied by the removal of the valence band holes and the subsequent
extinction of magnetism.
We predict that controlling the Hydrogen coverage can serve to tune a metal-insulator transition
and to achieve reversible switch of surface magnetism.

%
We are indebted to Prof.~P.~Esquinazi and Prof.~N.~Garc\'{\i}a for fruitful
discussions. We also acknowledge the unvaluable support of Dr.~C.~Francini
and Prof.~J.~H\"afner in the calculations using hybrid functionals.
This work was supported by the Spanish Ministry of Science and Technology 
(MAT2006-05122).


\begin{thebibliography}{99}
\bibitem{vasp1}
The PAW data sets with radial cutoffs of 1.3, 0.9, and 0.4~$\AA$ for Zn, O, and
H, respectively, were used with a plane-wave cutoff energy of 350 eV.
The HSE functional was built using a 0.38 fraction of 
exact exchange and a screening parameter of 0.2 $\AA^{-1}$.
\end{thebibliography}
\end{document}